\documentclass{iaus}
\usepackage{graphicx}
\title[Molecular Gas and Star Formation in Local Early-Type Galaxies] 
{Molecular Gas and Star Formation in Local Early-Type Galaxies}

\author[Bureau et al.]  
{M.\ Bureau,$^1$ T.\ A.\ Davis,$^1$ K.\ Alatalo,$^2$ A.\ F.\
  Crocker,$^3$ L.\ Blitz,$^2$ L.\ M.\ Young,$^4$ F.\ Combes,$^5$ M.\
  Bois,$^6$ F.\ Bournaud,$^7$ M.\ Cappellari,$^1$ R.\ L.\ Davies,$^1$
  P.\ T.\ de Zeeuw,$^{8,9}$ P.-A.\ Duc,$^{10}$ E.\ Emsellem,$^{8,6}$
  S.\ Khochfar,$^{11}$ D.\ Krajnovi\'{c},$^8$ H.\ Kuntschner,$^8$
  P.-Y.\ Lablanche,$^6$ R.\ M.\ McDermid,$^{12}$ R.\ Morganti,$^{13}$
  T.\ Naab,$^{14}$ T.\ Oosterloo,$^{13}$ M.\ Sarzi,$^{15}$ N.\
  Scott,$^1$ P.\ Serra$^{13}$ \and A.\ Weijmans$^{16}$
} \affiliation{$^1$University of Oxford, UK; $^2$University of
  California, Berkeley, USA; $^3$University of Massachusetts, Amherst,
  USA; $^4$New Mexico Tech, Socorro, USA; $^5$Observatoire de Paris,
  France; $^6$Observatoire de Lyon, France; $^7$CEA, Paris-Saclay,
  France; $^8$ESO, Garching, Germany; $^9$Leiden University, The
  Netherlands; $^{10}$Laboratoire AIM, Paris-Saclay, France;
  $^{11}$MPE, Garching, Germany; $^{12}$Gemini Observatory, Hilo, USA;
  $^{13}$ASTRON, Dwingeloo, The Netherlands; $^{14}$MPIA, Garching,
  Germany; $^{15}$University of Hertfordshire, Hatfield, UK;
  $^{16}$Dunlap Institute for Astronomy \& Astrophysics, University of
  Toronto, Canada}
\pubyear{2011}
\volume{277}  
\pagerange{1--4}
\jname{Tracing the Ancestry of Galaxies}
\editors{C. Carignan, K.C. Freeman \& F.. Combes, eds.}
\begin{document}
\maketitle
\begin{abstract}
  The molecular gas content of local early-type galaxies is
  constrained and discussed in relation to their evolution. First, as
  part of the ATLAS$^{\rm 3D}$ survey, we present the first complete,
  large ($260$~objects), volume-limited single-dish survey of CO in
  normal local early-type galaxies. We find a surprisingly high
  detection rate of $22\%$, independent of luminosity and at best
  weakly dependent on environment. Second, the extent of the molecular
  gas is constrained with CO synthesis imaging, and a variety of
  morphologies is revealed. The kinematics of the molecular gas and
  stars are often misaligned, implying an external gas origin in over
  a third of the systems, although this behaviour is drastically
  diffferent between field and cluster environments. Third, many
  objects appear to be in the process of forming regular kpc-size
  decoupled disks, and a star formation sequence can be sketched by
  piecing together multi-wavelength information on the molecular gas,
  current star formation, and young stars. Last, early-type galaxies
  do not seem to systematically obey all our usual prejudices
  regarding star formation, following the standard Schmidt-Kennicutt
  law but not the far infrared-radio correlation. This may suggest a
  greater diversity in star formation processes than observed in disk
  galaxies. Using multiple molecular tracers, we are thus starting to
  probe the physical conditions of the cold gas in early-types.
\end{abstract}
\firstsection 
%
\section{Introduction}
Early-type galaxies (ETGs), comprising ellipticals and lenticulars
(E/S0s), are generally considered ``red and dead'', with the bulk of
their stars homogeneously old, slightly less so at the lower masses
and in the field (e.g.\ \cite[Thomas et al.\ 2005]{tetal05}). This is
clearly an oversimplification, however, and residual star formation is
still present in at least $30\%$ of local ETGs. This is easily
quantified with near-ultraviolet imaging, particularly sensitive to
even small amounts of star formation (e.g.\ \cite[Yi et al.\
2005]{yetal05}), but is also clearly detected through optical
absorption lines (e.g.\ \cite[Kuntschner et al.\ 2010]{kuetal10}).

In this paper, we study the properties of the molecular gas in ETGs,
the fuel for star formation, through the $^{12}$CO(1-0) line. We focus
on the complete, volume-limited sample of $260$ local ETGs from the
ATLAS$^{\rm 3D}$ survey ($M_K<-21.5$~mag; \cite[Cappellari et al.\
2011]{caetal11}), as well as results from its predecessor ({\tt
  SAURON}; \cite[de Zeeuw et al.\ 2002]{zetal02}).
\section{Molecular gas properties}
In \cite[Young et al.\ (2011)]{yetal11}, we present a complete CO
single-dish survey of the ATLAS$^{\rm 3D}$ sample carried out with the
IRAM 30m telescope. With a sensitivity of $3$~mK $T_{\rm a}^*$ per
$30$~km~s$^{-1}$ channel ($\approx3\times10^7$~$M_\odot$ of molecular
hydrogen at the median distance), the main result is a surprisingly
high detection rate of $22\%$. Equally important, detections are
present throughout the so-called red sequence, formed of objects with
presumably low specific star formation rates. The molecular gas masses
span the range $10^7$--$10^9$~$M_\odot$, and both the detection rate
and the molecular-to-stellar mass ratios are independent of luminosity
and most structural and dynamical parameters probed. There is however
a clear dependence on the specific stellar angular momentum (as
quantified by the $\lambda_{\rm R}$ parameter; \cite[Emsellem et al.\
2007]{eetal07}), in the sense that slow-rotating galaxies are
deficient in molecular gas, and a possible weak dependence on
environment for fast-rotating galaxies.

\begin{figure}[t]
\begin{center}
  \includegraphics[width=0.575\textwidth]{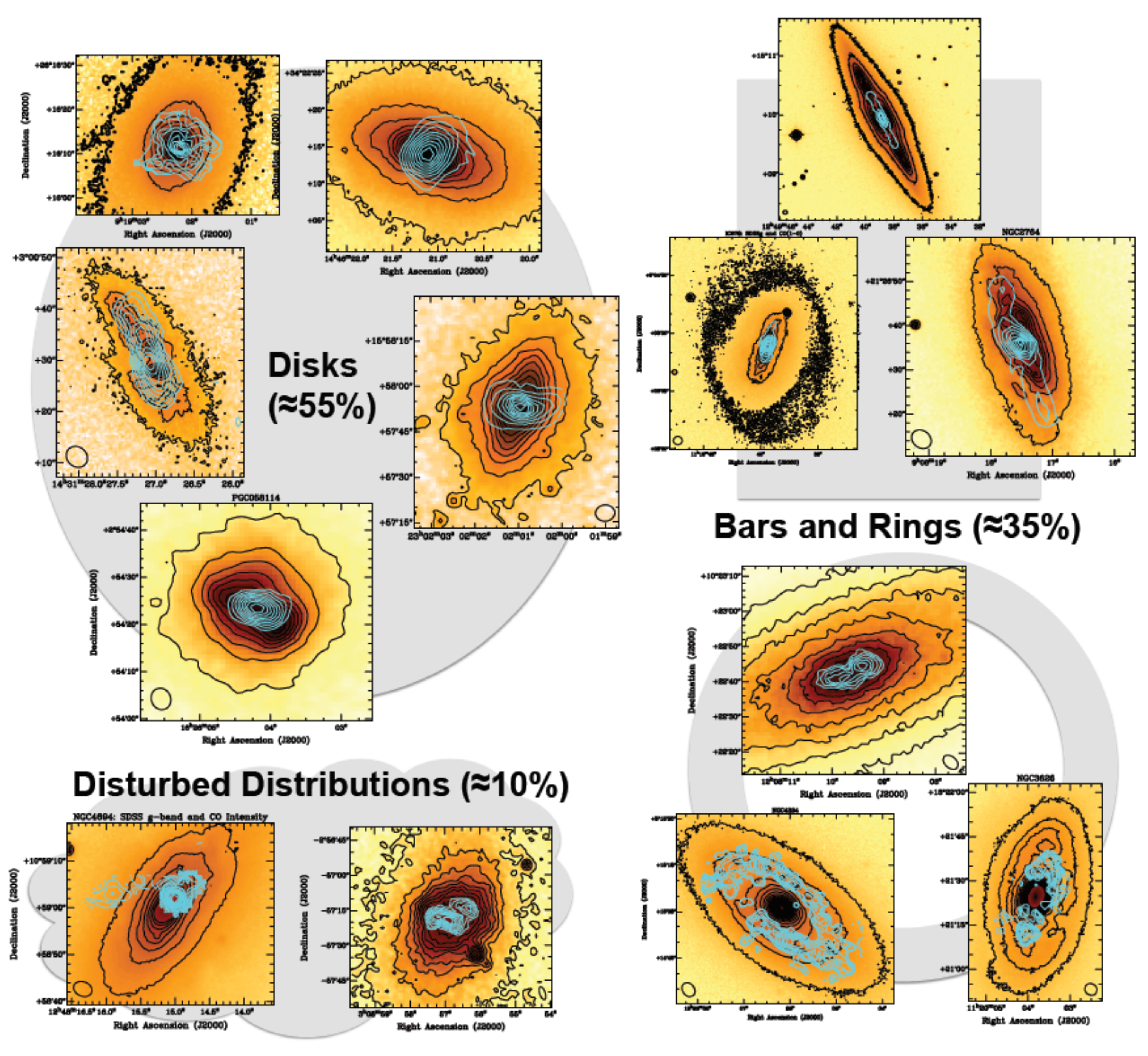}
  \caption{Mosaic showing CO distributions (solid blue lines) overlaid
    on optical images (colours and solid black lines) for selected
    sample galaxies. A range of morphologies is observed.}
 \label{fig:mosaic}
\end{center}
\end{figure}

We have strived to obtain follow-up interferometric line-imaging of
all CO detections, first with BIMA and PdBI, and now with CARMA
(Alatalo et al., in preparation). Figure~\ref{fig:mosaic} shows a
subset of the nearly $40$ galaxies observed so far, and it illustrates
the range of distributions detected. CO is generally
centrally-concentrated, with central disks, bars, rings, and
occasional disturbed morphologies. A careful comparison with the
spiral galaxies of the BIMA-SONG survey (\cite[Regan et al.\
2001]{retal01}) reveals that while the CO extent is generally smaller
in an absolute sense in ETGs, the extent distributions are identical
once scaled by a characteristic size (effective radius,
scalelength, isophotal diameter, etc).

\begin{figure}[t]
\begin{center}
  \includegraphics[width=0.4\textwidth]{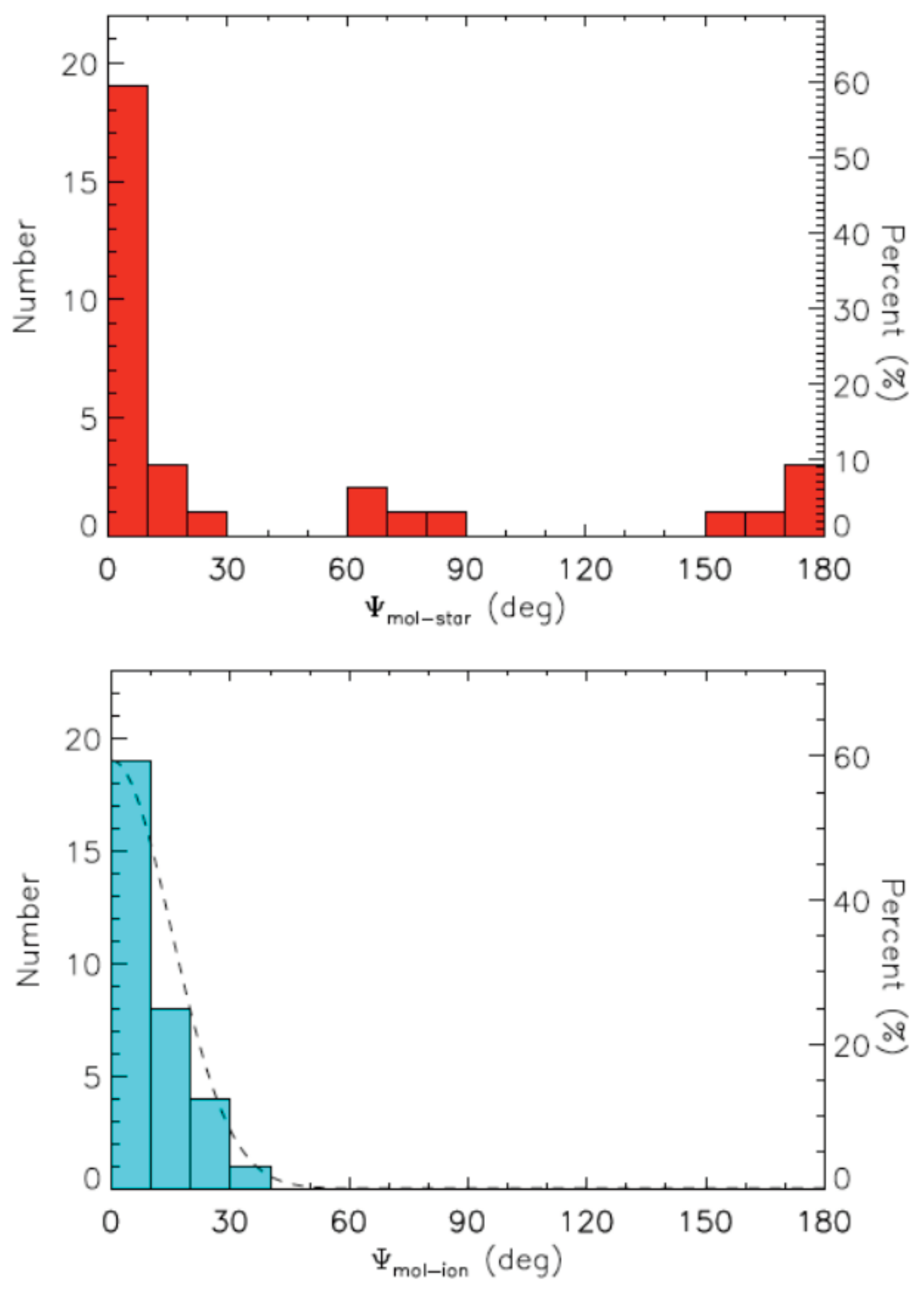}
  \includegraphics[width=0.45\textwidth]{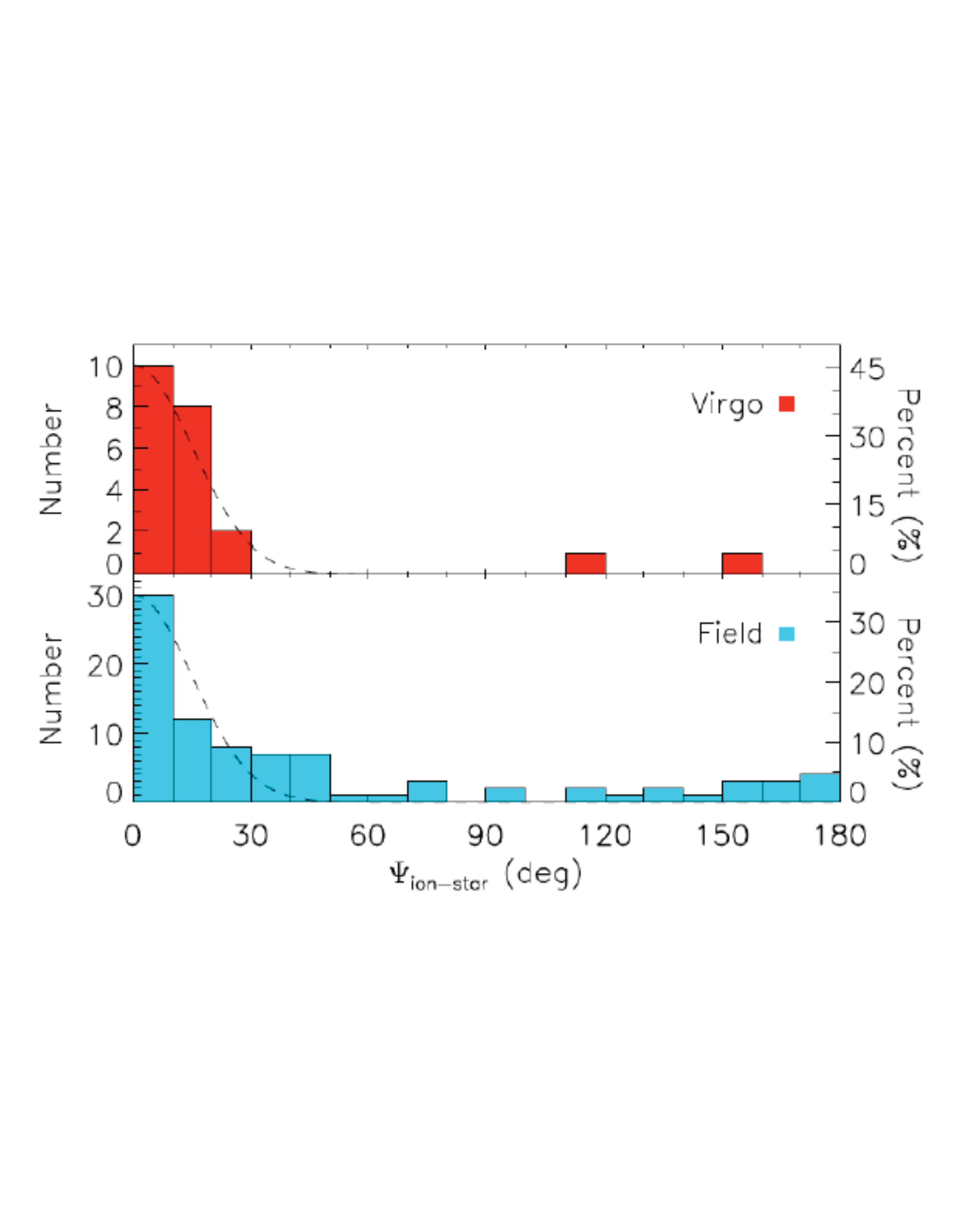}
  \caption{{\em Top-left:} Kinematic misalignment distribution between
    the stars and molecular gas. About $1/3$ of the objects are
    misaligned. {\em Bottom-left:} Kinematic misalignment distribution
    between the ionised and molecular gas. The two gas phases are
    always aligned. {\em Right:} Kinematic misalignment distribution
    between the ionised gas and stars, for fast-rotating galaxies in
    the Virgo cluster (top) and in the field (bottom). The galaxies
    are nearly always aligned in Virgo but show a range of
    misalignments in the field. Adapted from \cite[Davis et al.\
    (2011)]{detal11}.}
 \label{fig:kin}
\end{center}
\end{figure}

Taking full advantage of the three-dimensional nature of the {\tt
  SAURON} optical integral-field data {\em and} the CO synthesis
observations, we can compare the kinematics of the stars, ionised gas,
and molecular gas (Fig.~\ref{fig:kin}; \cite[Davis et al.\
2011]{detal11}). The kinematic major axes of the stars and molecular
gas are aligned in about $2/3$ of the CO-detected galaxies, implying
that the molecular gas is consistent with an internal origin (e.g.\
stellar mass loss) in $2/3$ of the cases overall, while it must have
an external origin (e.g.\ external accretion, minor merger) in at
least $1/3$ of the cases (some of the kinematically-aligned gas can
also have an external origin). Crucially, these statistics are
strongly dependent on environment. The stars and molecular gas are
nearly always kinematically-aligned in clusters (here Virgo), while
about half the galaxies are misaligned in the field. This implies that
external accretion of gas is very important in the field but is
possibly totally shut down in clusters. Interestingly, the ionised and
molecular gas are always kinematically-aligned, so that they must
share a common origin. As our ionised gas detection rate is much
higher, our conclusions are put on a much firmer basis by the ionised
gas.
\section{Star formation}
Comparing our molecular gas maps with the {\tt SAURON} data in a
spatially-resolved way, the molecular gas appears to be forming young
central disks in a number of objects (\cite[Crocker et al.\
2011]{cretal11}). This is revealed by fast-rotating decoupled central
stellar components, depressed stellar velocity dispersions, ionised
gas emission with [O{\small III}]/H$\beta$ line ratios typical of star
formation, enhanced H$\beta$ absorption, etc. However, this is not
always the case. Molecular gas is sometimes associated with [O{\small
  III}]/H$\beta$ line ratios more typical of those expected from
evolved stellar populations, and occasionally no sign of a young
stellar population is detected at all.

\begin{figure}[t]
\begin{center}
  \includegraphics[width=0.6\textwidth]{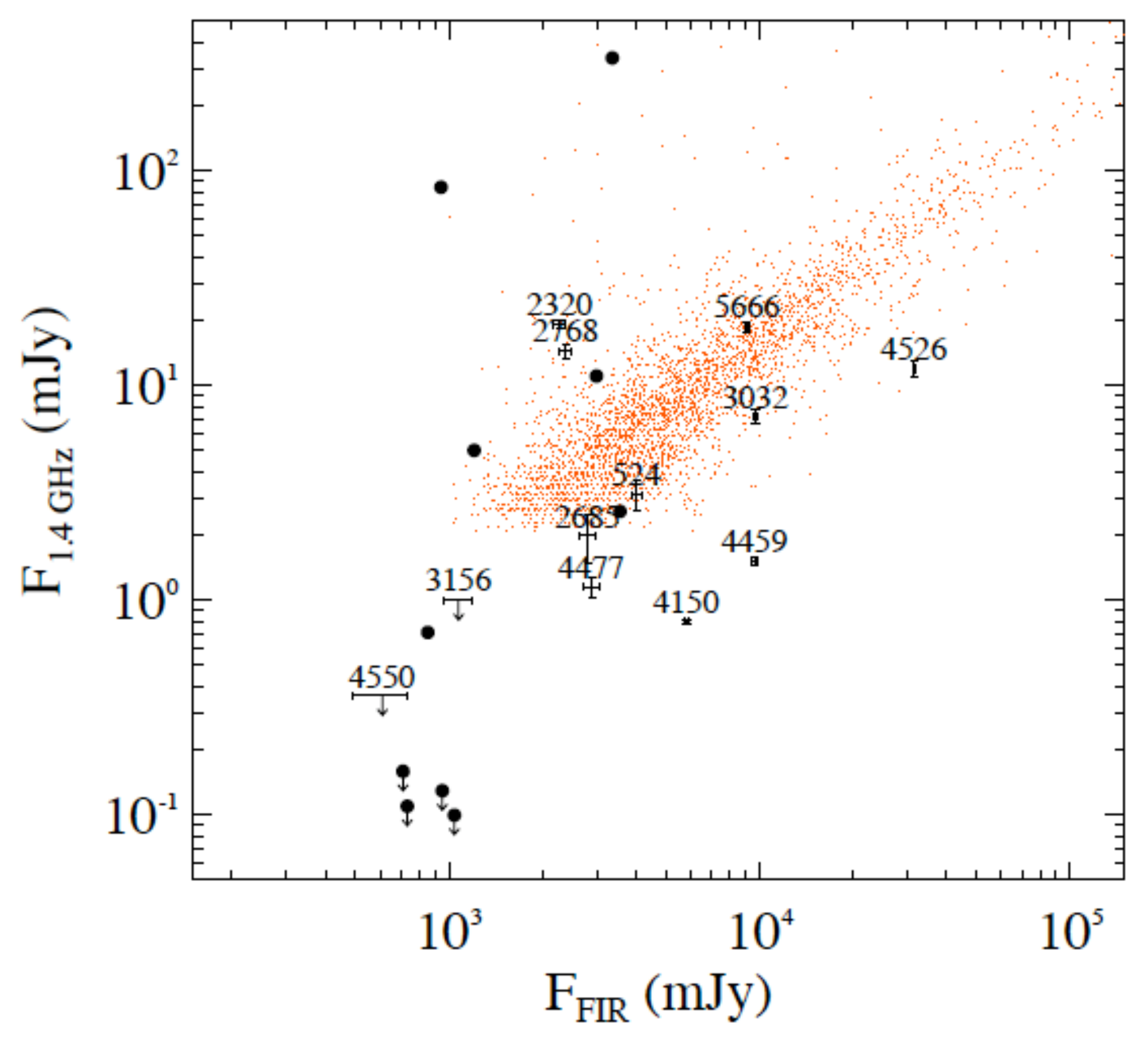}
  \caption{Far infrared-radio continuum ($1.4$~GHz) correlation of the
    {\tt SAURON} ETGs (black dots) compared to that of the galaxies in
    the Uppsala General Catalogue (red dots; mostly disk
    galaxies). Black dots with error bars are CO-detected and
    preferentially lie on the FIR-bright side of the relation. Adapted
    from \cite[Crocker et al.\ (2011)]{cretal11}.}
 \label{fig:FIR-radio}
\end{center}
\end{figure}

As ETGs are also (on average) significantly different dynamically from
disk galaxies (e.g.\ $Q$ parameter), it is clear that they represent a
different and unique environment in which to probe star formation,
both its usual tracers and possible causal relations. After carefully
accounting for internal extinction, we find that the Kennicutt-Schmidt
relation and a constant star formation efficiency (relating the
surface density of molecular gas to that of star formation) are both
consistent with the data. However, the far infrared (FIR)-radio
continuum correlation is not satisfied, with too many FIR-bright
galaxies (Fig.~\ref{fig:FIR-radio}), and different infrared star
formation tracers do not agree, perhaps implying significant dust
heating from non-star formation sources.

Understanding possible differences in the star formation processes of
ETGs requires a good knowledge of the physical conditions in the
interstellar medium. We have thus started a programme to constrain the
opacity, column density, volume density, temperature, and other
parameters of the cold gas in our sample galaxies, using multiple
molecular line tracers (e.g.\ $^{13}$CO, HCN, HCO$^+$). Preliminary
results indicate occasionally enhanced $^{13}$CO and suppressed
HCO$^+$ (\cite[Krips et al.\ 2010]{kretal10}; Crocker et al., in
preparation).

Observing molecular tracers other than $^{12}$CO is challenging with
the current generation of instruments, but with its vastly improved
sensitivity and angular resolution, ALMA will both broaden and increase
the pace of discoveries. It will also allow to probe trends as a
function of both lookback time and environment. Herschel will
similarly improve our understanding of the gas (atomic, ionised and
molecular) and dust associated with the molecular gas, in particular
the dust-to-gas ratio.
%
%

%
\end{document}